\documentclass[prd,superscriptaddress,nofootinbib]{revtex4}
\usepackage{amssymb}
\usepackage{amsmath}
\usepackage[]{hyperref}
\usepackage{cleveref}
\usepackage{graphicx}

\Crefname{equation}{Eq.}{Eqs.}
\Crefname{figure}{Fig.}{Figs.}
\Crefname{section}{Sec.}{Secs.}
\Crefrangeformat{equation}{Eqs.~(#3#1#4)--(#5#2#6)}

\numberwithin{equation}{section}

\begin{document}

\title{PeV IceCube signals and $H_0$ tension in the framework of Non-Local Gravity}

\author{Salvatore Capozziello}
\email{capozziello@na.infn.it}
\affiliation{Dipartimento di Fisica ``E. Pancini", Universit\`a di Napoli ``Federico II", Via Cinthia 9, I-80126 Napoli, Italy.}
\affiliation{Scuola Superiore Meridionale, Largo S. Marcellino 10, I-80138 Napoli, Italy.}
\affiliation{Istituto Nazionale di Fisica Nucleare (INFN), Sez. di Napoli, Via Cinthia 9, I-80126 Napoli, Italy.}

\author{Gaetano Lambiase}
\email{lambiase@sa.infn.it}
\affiliation{Dipartimento di Fisica "E.R. Caianiello", Universit\`a di Salerno, Via Giovanni Paolo II 132, I-84084 Fisciano (SA), Italy.}
\affiliation{Istituto Nazionale di Fisica Nucleare (INFN), Sez. di Napoli, Gruppo Collegato di Salerno, Italy.}

\begin{abstract}
We study possible effects  of non-local gravity corrections on the recent discovery by the IceCube collaboration reporting high-energy  neutrino flux  detected at energies of  order PeV. Considering the 4-dimensional operator $\sim y_{\alpha \chi}\overline{{L_{{\alpha}}}}\, H\, \chi$, it is possible to  explain both the IceCube neutrino rate and the   abundance of Dark Matter,  provided that non-local corrections are present in the cosmological background.  Furthermore, the mechanism could constitute a natural way to address the $H_0$ tension issue.
\end{abstract}

\maketitle

\section{Introduction}

General Relativity (GR) provides  the  best theory to describe  gravitational interaction confirmed by  experiments and  observations ranging from  Solar System up to cosmology.  Despite these successes, GR is affected by several shortcomings, arising both at ultraviolet (UV) and infrared (IR) scales. These suggest that GR is an effective theory not working at any energy scale and, therefore, it could be not the final theory of gravity.
 At fundamental level,    space-time singularities emerge and the theory is  inconsistent  with a quantum field formulation. At  astrophysical scales,  discrepancies between  theoretical  and  observed dynamics  in galaxies and clusters of galaxies  lead to the missing matter problem. In general, this issue is addressed by   introducing Dark Matter (DM), which accounts for more or less 25\% of the Universe content.  At cosmological scales,  current observations show that the present Universe is in an accelerating phase. This can be explained by means of  the so-called Dark Energy (DE).
However,  at the moment, there is no final evidence for particles addressing  DM and DE at fundamental scales so  modifying  the gravitational sector is becoming a realistic approach to the dark side problem (see, e.g. \cite{Capozziello:2011et}).

A possibility to overcome  GR shortcomings, both at UV and IR scales, is  considering non-local theories of gravity   which are characterized by  actions of the form
\begin{equation}\label{NL:22}
\mathcal{S}=\int d^4x \sqrt{-g}\,R\left[1+F(\Box^{-1}R)\right]\,,
\end{equation}
where $R$ is the Ricci curvature scalar and $\Box$ the d'Alembert operator. From a fundamental physics  point of view,  non-local theories emerges in view to obtain  renormalizable and unitary effective gravity models \cite{Biswas:2011ar, Anupam}.
Remarkably, the term $\square^{-1}R$ can account for the late-time cosmic expansion without invoking any DE \cite{Deser:2007jk}.   

In \cite{SalvPLB22, Capozziello:2022lic}, it is shown that non-local terms are related to conserved quantities coming from the existence of Noether symmetries which select the form of the function $F(\square^{-1}R)$.  Specifically, if a Noether symmetry exists, the Lagrangian in \eqref{NL:22} can result of the form
\begin{equation}
\label{Noether}
{\cal L}=R\left(c_1+c_2   \, e^{\frac{3 q-1}{3 (2 q-1)} \Box^{-1}R}\right)\,,
\end{equation}
where $c_{1,2}$ are gravitational couplings, $q$ is a free parameter related to the symmetry.  Here the case $q=1/2$ has to be excluded and this fact will be crucial for the discussion below. In a Friedman-Lema\^itre-Roberson-Walker cosmology, dynamics is exactly solved by 
\begin{eqnarray}
&&a(t)=a_0\, t^q \label{pow}\\
&&R(t)=6q\left(\frac{1-2q}{t^2}\right)  \\
&&\Box^{-1}R=\Lambda-\frac{6 q (2 q-1) \log(t)}{3 q-1} \\
\end{eqnarray}
We can recast the evolution of the scale factor of the Universe \eqref{pow} as 
\begin{equation}\label{powerlaw}
   a(t) = a_*  \left(\frac{t}{t_*}\right)^q,
 \end{equation}
where  the index $*$ points out the instant $t_*$ (and the corresponding temperature $T_*$) at which the Universe starts to expand.
The expansion rate  $H={\dot a}/a$, where  dot indicates the  derivative with respect to the cosmic time $t$, can be cast as  \cite{BD}
 \begin{equation}\label{H=AHGRIce}
   H_{non-local}(T)=A(T)H_{GR}(T)\,,
 \end{equation}
where $H_{GR}$ is the expansion rate  in GR ($H_{GR}=\sqrt{\frac{8\pi \rho}{3M_P^2}}=\sqrt{\frac{8\pi^3 g_*}{45}}\, \frac{T^2}{M_P}$), and $A(T)$ is an amplification factor.  Using the power law solution and the conservation of  entropy ($Ta=T_* a_* = T_0$, with $T_0$ the temperature of the present Universe at $a_0=1$), one can parameterize the amplification factor as
 \begin{equation}\label{A(T)Ice}
   A(T) =\eta \left(\frac{T}{T_*}\right)^\nu , \qquad
   \nu = \frac{1}{q}-2\,, \qquad \eta = 2q\,,
 \end{equation}
where $\{\eta, \nu\}$ characterize the given cosmological model\footnote{Investigations   have been performed in different cosmological scenarios \cite{fornengoST,BD,gondolo}, where the parameter $\nu$ labels the cosmological model. It is $\nu=2$ in Randall-Sundrum type II brane cosmology \cite{randal}; $\nu=1$ in kination models \cite{kination}; $\nu=0$ in cosmologies with an overall boost of the Hubble expansion rate \cite{fornengoST}; $\nu=-0.8$ in scalar-tensor cosmology \cite{fornengoST}; $\nu=2/n-2$ in $f(R)$ gravity with $f(R)=R+\alpha R^n$ \cite{lamb}.}.
To preserve the successful predictions of Big Bang Nucleosynthesis (BBN), one refers to the pre-BBN epoch  which is  not directly constrained by cosmological observations \cite{fornengoST,BD}.
Therefore, $A(T) \neq 1$ at early time $(T\gtrsim T_* \gg T_{BBN}$), and $A(T)\to 1$ at $T=T_*$ (before BBN begins).

We want to show that non-local gravity (\ref{NL:22}), in particular the  model \eqref{Noether} selected by Noether symmetries,    allows to get a consistent explanation of PeV-DM and IceCube neutrinos events reported by   the IceCube Collaboration \cite{2a}   with energies $\sim 1 $~PeV \cite{2a}.

Candidates for the generation of such  high energy neutrino events are various astrophysical sources,
although there is  no clear correlations with the known astrophysical hot-spots like the  supernova remnants  or active galactic nuclei  \cite{Aartsen:2016oji}. This suggests the possibility that neutrinos could arise from the decay of PeV-DM particles \cite{merle,lambIce}.
To explain the PeV-DM relic density and the decay rate required for IceCube, one can take into account   the minimal DM-neutrino (4-dimensional) interaction
 \begin{equation}\label{Ldim4}
   {\cal L}^{d=4} = y_\alpha \, \bar L_\alpha \cdot H \chi\,, \qquad \alpha = e, \mu, \tau\,,
 \end{equation}
where $\alpha $ indicates the  mass eigenstates of the three active neutrinos, $\chi$ the DM particle, $H$ the Higgs doublet, $L_\alpha$ the left-handed lepton doublet, and $y_{\alpha \chi}$ the Yukawa couplings. Notice that the 4-dimensional operator (\ref{Ldim4}) fails to explain both  IceCube data and DM relic abundance, if the cosmological background evolves according to the standard Einstein-Friedman field equations \cite{merle,lambIce}. 
In the latter case, in fact, one finds that the relic abundance (induced by inverse decay) is \cite{merle}
 \begin{equation}\label{DMinvdec}
   \Omega_{DM}h^2 =0.1188 \left(\frac{106.75}{g_*}\right)^{3/2}\frac{\sum_\alpha |y_{\alpha \chi}|^2}{7.5\times 10^{-25}}\,.
 \end{equation}
From (\ref{DMinvdec}), it follows that in order  to get the correct relic  DM  abundance ($\Omega_{DM}h^2 \sim 0.1188$), one needs
 \begin{equation}\label{ychi}
   \sum_{\alpha=e, \mu,\tau}|y_{\alpha\chi}|^2=7.5\times 10^{-25}\,.
 \end{equation}
However, Eq. (\ref{DMinvdec}) is incompatible with the value of $\sum_{\alpha=e, \mu,\tau}|y_{\alpha\chi}|^2$ needed to explain the IceCube data. To fix this point, let us  note that the DM lifetime $\tau_\chi=\Gamma_\chi^{-1}$ has to be larger that the age of the Universe, $\tau_\chi> t_U\simeq 4.35\times 10^{17}$sec. Moreover, IceCube spectrum sets  constraints on lower bounds of DM lifetime $\tau_\chi^b\simeq 10^{28}$sec, i.e. $\tau_\chi \gtrsim \tau_\chi^b$, which is (approximatively) model-independent (see \cite{merle}). From (\ref{ychi}), one obtains $ \Gamma_\chi\simeq 4.5 \times 10^4 \frac{m_\chi}{\text{1PeV}}\text{sec}^{-1}$, that is $ \tau_\chi \simeq 2.2\times 10^{-5} \frac{\text{1PeV}}{m_\chi}\text{sec}\ll t_U$.
Observations of IceCube collaboration require, however, that the DM decay lifetime has to be $\tau_\chi \sim 10^{28}$ sec, which implies
 \begin{equation}\label{yIceCube}
   \sum_{\alpha}| y_{\alpha\chi}|^2\simeq 10^{-58}\,,
 \end{equation}
which is $\sim 33$ order of magnitudes smaller than the value of $\sum_{\alpha=e,\mu, \tau}|y_{\alpha\chi}|^2\sim 10^{-25}$ needed to explain the DM relic abundance, see (\ref{ychi}). As a consequence, the IceCube high energy events and the DM relic abundance are not compatible with the DM production, if the latter is ascribed to the 4-dimensional operator $\overline{ L_{\alpha}}H\chi$ and it is assumed that the evolution of the Universe is governed by GR. Such a discrepancy can be  avoided if  non-local  gravity corrections are assumed in the cosmological evolution as we will show below.

\section{PeV neutrinos and DM relic abundance in non-local gravity}
\setcounter{equation}{0}

Let us consider  a   {\it freeze-in} production where  DM particles are never in thermal equilibrium because they interact very weakly and are gradually produced from the hot thermal bath. This occurs owing to a feeble coupling to  the Standard Model particles (at $T\gg m_\chi$) allowing the DM particles to remain in the observed Universe due to  small  back-reaction rates and to  slow  decay processes. Therefore a sizable DM abundance is allowed, at least, until the temperature falls down to $T\sim m_\chi$.  Temperatures below $m_\chi$ are such that the phase-space of DM particles  is kinematically difficult to access \cite{hall,merle}.

\begin{figure}[!t]
  \centering
  \includegraphics[width=5.8cm]{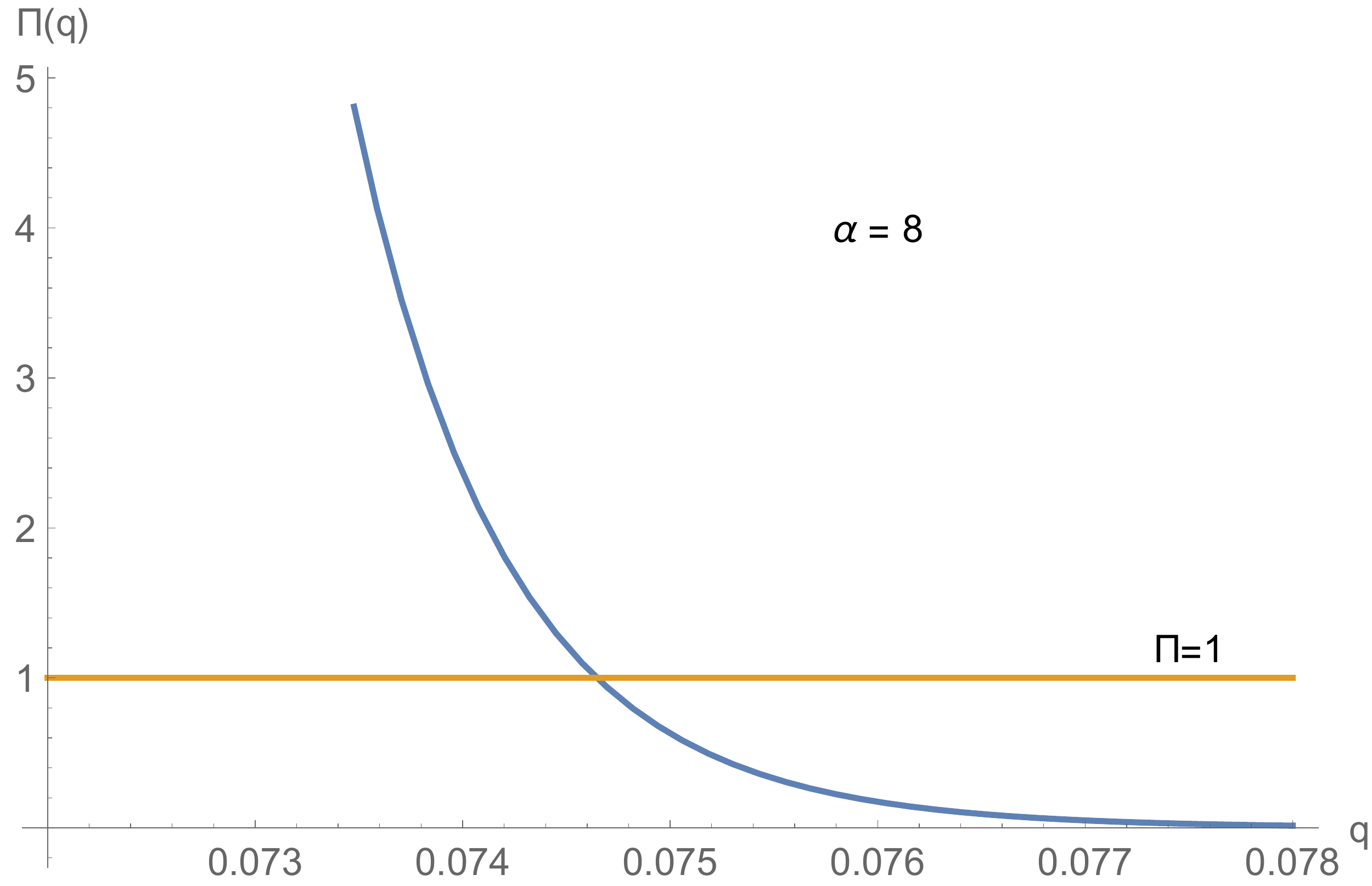}\qquad \qquad
  \includegraphics[width=5.8cm]{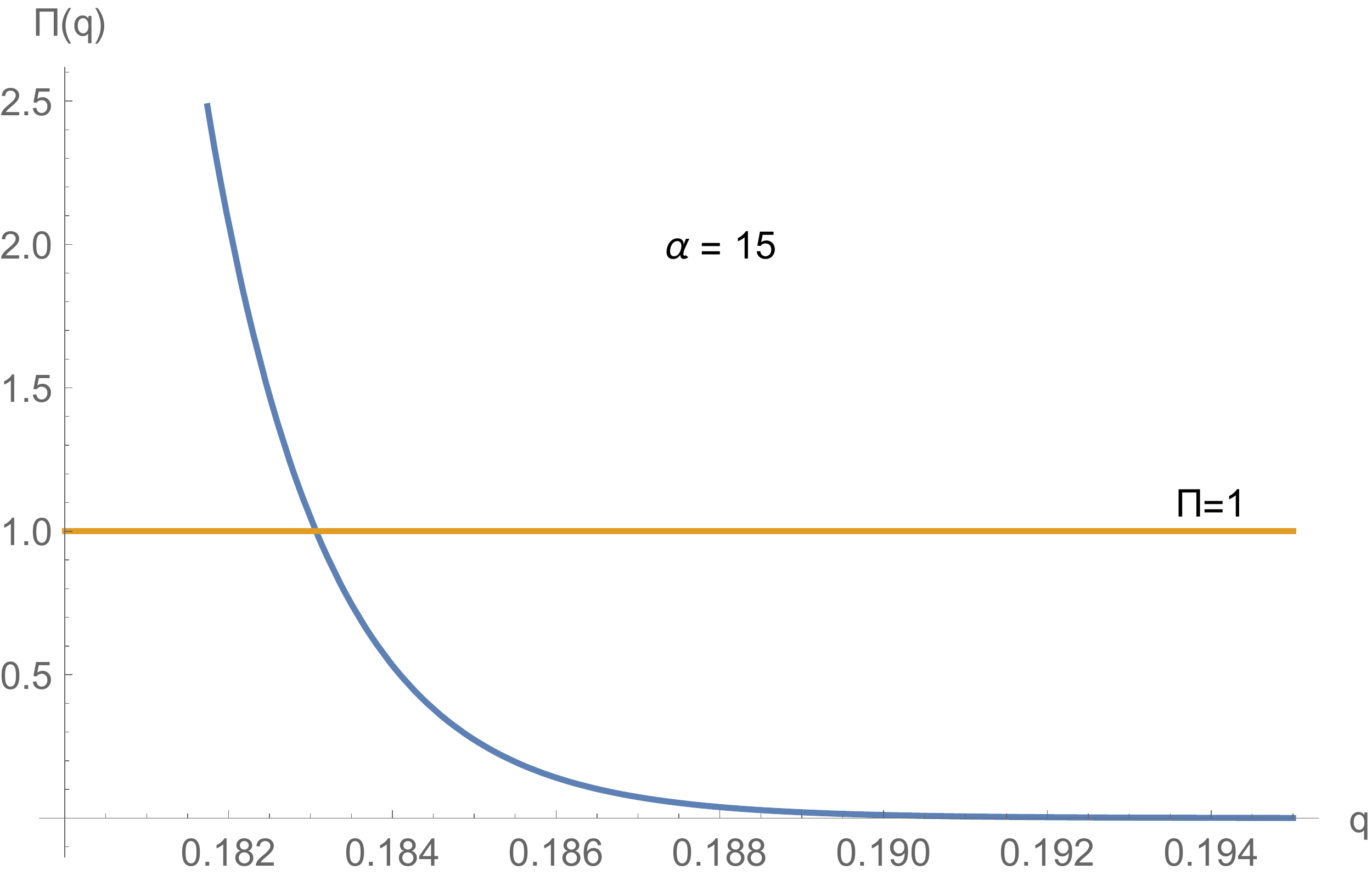}\\
  \caption{$\Pi$ vs $q$ for $\alpha=8,15$.}\label{alpha15}
\end{figure}

The evolution of  DM particles is governed by the Boltzmann equation. Let us denote with $Y_\chi=n_\chi/s$ the DM abundance, where $n_\chi$ is the number density of the DM particles and $s=\frac{2\pi^2}{45}g_*(T)T^3$ the entropy density ($g_*$ denotes the degrees of freedom). By using the Boltzmann equation
and assuming that the relativistic degrees of freedom are constant, i.e. $dg_*/dT=0$, the DM relic abundance can be cast in the form
 \begin{equation}\label{DM1}
   \Omega_{DM}h^2=\frac{2 m_\chi^2 s_0 h^2}{\rho_{cr}}
   \int_0^\infty\frac{dx}{x^2}\left(-\frac{dY_\chi}{dT}\Big|_{T=\frac{m_\chi}{x}}\right)\,,
 \end{equation}
where $x=m_\chi/T$. Here $s_0=\frac{2\pi^2}{45}g_*T_0^3\simeq 2891.2/$cm$^3$ is the present value of the entropy density, and $\rho_{cr}=1.054\times 10^{-5}h^2$GeV/cm$^3$ the critical density.
In the case of  4-dimensional operator (\ref{Ldim4}), the dominant contributions to DM production is given by the {\it inverse decay} processes $\nu_\alpha+H^0\to \chi$ and $l_\alpha + H^+\to \chi$  occurring when $m_\chi> m_H+m_{\nu, l}$. The interaction rate  is  \cite{merle}
\begin{equation}\label{intrate}
\Gamma_\chi=\sum_\alpha \frac{|y_{\alpha\chi}|^2}{8\pi}\, m_\chi\,, \quad  \alpha=e, \mu, \tau\,.
\end{equation}
Therefore, according to (\ref{A(T)Ice}), the inverse decay processes turn out to be
\begin{equation}\label{dYMC}
\frac{dY_\chi}{dT} = -\frac{m_\chi^2 \Gamma_\chi}{\pi^2 s H_{non-local}}\, K_1\left(\frac{m_\chi}{T}\right)\,,
\end{equation}
where $K_1(x)$ is a modified Bessell function of  second kind. From (\ref{H=AHGRIce}) and (\ref{A(T)Ice}), it is
 \begin{equation}\label{Heta}
  s H_{non-local}=\frac{3.32\pi^2}{45}g_*^{3/2}\eta\left(\frac{m_\chi}{T_*}\right)^\nu \frac{1}{x^{5-\nu}}\,,
  \quad x\equiv \frac{T}{T_*}\,.
 \end{equation}
By inserting (\ref{dYMC}) into (\ref{DM1}), one obtains the DM relic abundance
 \begin{eqnarray}
   \Omega_{DM}h^2 &=& \frac{45h^2}{1.66\pi^2 g^{3/2}}\frac{s_0 M_{Pl}}{\rho_{cr}}\frac{\Gamma_\chi}{m_\chi}\frac{2^{2+\nu}}{\eta}\left(\frac{T_*}{m_\chi}\right)^\nu
   \Gamma\left(\frac{5+\nu}{2}\right)\Gamma\left(\frac{3+\nu}{2}\right) \label{DMeta} \\
  &\simeq & 0.1188 \left(\frac{106,7}{g_*}\right)^{3/2}\frac{\sum_\alpha |y_{\alpha\chi}|^2}{10^{-58}}\, \Pi\,,
  \end{eqnarray}
where $\Gamma(x)$ is the Gamma function, and
 \begin{equation}\label{Phi}
  \Pi\equiv 10^{-33}\, \frac{2^{\nu}}{7.5 \eta}\left(\frac{T_*}{m_\chi}\right)^\nu
   \frac{\Gamma\left(\frac{5+\nu}{2}\right)\Gamma\left(\frac{3+\nu}{2}\right)}{\Gamma\left(\frac{5}{2}\right)\Gamma\left(\frac{3}{2}\right)}\,.
 \end{equation}
The function $\Pi$  depends on  non-local gravity parameters  $\nu$ and $\eta$, that is $q$.
The DM relic abundance (requiring $\Omega_{DM}h^2 \sim 0.1188$ \cite{Planck}) and the IceCube data (requiring $\sum_\alpha |y_{\alpha\chi}|^2 \sim 10^{-58}$ \cite{2a}) can be consistently explained provided $\Pi\simeq 1$. To search the values of $q$ such that the condition $\Pi\sim 1$ is fulfilled, we parameterize the temperature $T$ as
\begin{equation}\label{Talpha}
  T= 10^\alpha \text{GeV}\,,
\end{equation}
with $\alpha > 6$, that means we consider temperature greater than the DM mass, $T> m_\chi$ (and therefore greater than the BBN temperature $T_{BBN}\sim 1$MeV).
In Fig. \ref{alpha15}, we report $\Pi$ vs $q$ for fixed values of the parameter $\alpha$. The latter ranges from $\alpha=8$ to $\alpha=15$ to account for  scales  from DM  to GUT. As we can see, the parameter $q$ varies in the range $0.07 \lesssim q \lesssim 0.18$. 
This is the result  we want. It shows that, for $q<1/2$,  the expansion rate of the
Universe is modified,  as well as the Boltzmann equations and any contribution to the energy density due to  matter and geometry sectors. The effect on the  evolution makes the  
relic DM abundance and  IceCube (1PeV)  neutrino signals  consistent with the 
observations. It is worth noticing that the presence of non-local term in the Lagrangian \eqref{Noether} exclude $q=1/2$, corresponding to the radiation dominated epoch  of the standard cosmological model. It is worth noticing that the presence of the operator $\Box^{-1}$ explains the current late-time accelerated cosmic expansion without invoking any Dark
Energy. Specifically, it enables a delayed response to the radiation-matter transition which could explain
the current cosmic acceleration \cite{Deser:2007jk}.
Such a result  opens  possibilities to probe new physics beyond GR and it is a natural way to potentially address the $H_0$ tension issue \cite{Tension}.

\section{Conclusions}
To reconcile the current bound on DM relic abundance with IceCube data, in terms of the 4-dimensional operator ${\cal L}_{d=4}=y_{\alpha \chi}\overline{{L_{{\alpha}}}}\, H\, \chi$, we  considered cosmological solutions coming from  non-local gravity.   Specifically, we  focused on cosmological power law solutions emerging from a non-local model selected by the presence of Noether symmetries. These exact solutions allow to account for  IceCube observations (high energy neutrinos) and  DM relic abundance, observed  in a minimal particle physics model. The main idea relies on the fact that, in non-local cosmology,  the expansion rate of the Universe can be cast in the form $H(T)=A(T)H_{GR}(T)$, encoding in $A(T)$ the parameters characterizing the cosmological  model. As a consequence, also the thermal history of particles results  modified. On the other hand,  the  PeV signals reported by the IceCube collaboration could be a straightforward signature for non-local effects emerging at cosmological level. Finally, the stretching induced by non-local term in the Hubble parameter, led by the function $A(T)$, could  solve, in principle,  the $H_0$ tension in the framework of fundamental physics. In a forthcoming paper, we will discuss in detail this topic from an observational point of view.

\section*{Acknowledgments }
The authors acknowledge the support of  Istituto Nazionale di Fisica Nucleare (INFN), Sez. di Napoli, {\it iniziativa specifiche} QGSKY and MoonLIGHT2.

\section*{Data Availability  Statement }
Data sharing not applicable to this article as no datasets were generated or analysed during the current study.



\end{document}